\documentclass[aps,amsmath,prl,twocolumn,a4paper,showpacs,floatfix]{revtex4}
\usepackage{graphicx}

\newcommand{\beq}{\begin{equation}}
\newcommand{\eneq}{\end{equation}}
\newcommand{\bea}{\begin{eqnarray}} 
\newcommand{\enea}{\end{eqnarray}}
\newcommand{\bean}{\begin{eqnarray*}}
\newcommand{\eean}{\end{eqnarray*}}
\newcommand{\pp}{\mathbf{p}}

\begin{document}

\title{Rashba-control for the spin excitation of a fully spin polarized  vertical quantum dot}

\author {P.Lucignano$^{1,2}$} \author{B.Jouault$^3$}
\author{A.Tagliacozzo$^{1,2}$}\author{B.L.Altshuler$^{4}$}

\affiliation{$^1$ Coherentia-INFM,  
Monte S.Angelo - via Cintia, I-80126 Napoli, Italy }
\affiliation{$^2$ Dipartimento di Scienze Fisiche Universit\`a di 
         "Federico II " Napoli, Napoli Italy}
\affiliation{$^3$ GES, UMR 5650, Universit\'e Montpellier $II$,
         34095 Montpellier Cedex 5, France}
\affiliation{$^4$ Phisics department, Princeton University, Princeton, NJ08540, USA and \\NEC Research Institute, Princeton, NJ08540, USA}
		 
\date{\today}
\begin{abstract}
Far infrared radiation absorption of a quantum dot with few electrons
in an orthogonal magnetic field could monitor the crossover to the
fully spin polarized state.  A Rashba spin-orbit coupling can tune the
energy and the spin density of the first excited state which has a
spin texture carrying one extra unit of angular momentum.  The spin
orbit coupling can squeeze a flipped spin density at the center of the
dot and can increase the gap in the spectrum.
\end{abstract}

\maketitle

\textit{Introduction}.
Quantum dots (QD's), confining one or few active electrons
\cite{kouwenhoven}, have been proposed as devices for the future
quantum electronics. One of the possibilities is to operate on the
spin of the trapped electrons as a qubit \cite{divincenzo}. In a
different proposal the QD controls the nuclear spins embedded in the
crystal matrix via hyperfine coupling\cite{kane}. In both cases the
polarization of the spins is expected to last long enough at low
temperatures, so that the quantum computation can be carried
out. Controlled spin transfer between electrons and nuclei has been
demonstrated to be possible in a spin polarized two-dimensional
electron gas (2DEG) \cite{smet}.  In a 2DEG fully spin polarized
quantum Hall states are used to manipulate the orientation of nuclear
spins.  Low lying skyrmion states at filling close to one are used to
reset the nuclear spin system by inducing fast spin relaxation. In the
presence of a magnetic field $B$ orthogonal to the dot, the relaxation
mechanism seems to be dominated by hyperfine interaction for $B<0.5 T$
and by spin-orbit (SO) coupling assisted by phonons for higher fields
\cite{khaetskii}.

The Rashba SO interaction \cite{rashba}, which arises in QD structures
from the lack of inversion symmetry caused by the two-dimensional (2D)
confinement, can be controlled by gate voltages parallel to the $x-y$
structure \cite{nitta}.  This possibility has been beautifully shown
in $InGa As -$based 2DEG\cite{meijer} and in a recent experiment on
large lateral QD, where the conductance has been tuned from the weak
localization limit, without SO coupling, to the antilocalization
limit, with SO\cite{miller}.  The inverse relaxation time $1/T_1$ has
also been probed recently by transport across a single QD
\cite{huettel}.

Thanks to the combined effect of $e-e$ interactions and of an
appropriate $B>B^*$ orthogonal to the dot ($\parallel \hat z $), the
QD becomes a `maximum density droplet' (MDD) with a fully spin
polarized (FSP) ground state (GS)\cite{osterk}.  In this paper we show
that controlled SO interaction of a FSP QD, made out of III-V
semiconductors, can be used to adiabatically modify the low lying
energy states of few ($N=2,3,4$) trapped electrons and their spin
density.  The Rashba-SO coupling contributes to a well defined
collective spin excitation (first excited state (FES)), with a change
of the spin density, w.r.to the GS, localized at the origin of the
QD. This excitation can be pumped with Far Infrared Radiation (FIR).
Using numerical diagonalization for few electrons in the dot, we find
that the absorption intensity for circularly polarized FIR is strongly
enhanced when the crossover to the FSP state is completed (see
Fig.\ref{ap}). The spin density can be squeezed at the center of the
dot by increasing the SO coupling $\alpha$ (see Fig.\ref{squeez}).
Meanwhile the gap between the GS and FES increases with $\alpha$ (as
shown in Fig.\ref{plot3}d) for N=3).

\textit{QD  spin properties tuned   by  SO for  $B>B^*$}.
The electrons are confined in two dimensions $(x,y)$ by a parabolic
potential of characteristic frequency $\omega _d $, in the presence of
$\vec B = -B \hat z$. The single particle hamiltonian for the $i-$th
electron, in the effective mass approximation ($m_e^*$), is:
\beq
H_{(i)} = \frac{1}{2m_e^*} \left(\vec p_i + \frac{e}{c} \vec A_i
\right)^2+\frac{1}{2} m_e^* \omega_d^2 \vec r _i ^2\;\;,\label{hamsp}
\eneq
with $\vec A_i=B/2(y_i,-x_i,0)$, and $-e$ is the electron charge.

The Zeeman spin splitting would only mask our results with additional
inessential complication and we neglect it\cite{macdonald}. The single
particle Darwin-Fock states $\phi _{nm}$ are the eigenfunctions of the
2D harmonic oscillator with frequency $\omega _o = \sqrt { \omega _d^2
+{\omega _c^2}/{4}}$, where 
$\omega _c = eB/m_e^*c$, the cyclotron frequency.  $m$ is the angular
momentum in the $z$ direction ( $m \in (-n,-n+2,...,n-2,n)$ with $n
\in (0,1,2,3,...)$ ).  The radial size of the $\phi _{nm}$'s is $ \sim
l=\sqrt{\hbar / m_e^*\omega_o}$, the characteristic length due to the
the lateral geometrical confinement in the dot, inclusive of the $B$
field effects.
  
The corresponding single particle energy levels are: $\epsilon _{n ,m}
= (n +1) \hbar \omega _o - m \hbar \omega_c/2$.  In the
absence of SO, the full hamiltonian for the dot, inclusive of the
Coulomb interaction between the electrons (parametrized by $U$) is: $
H=\sum_{i=1}^{N} H_{(i)}+ \sum_{{i<j} \atop { i,j=1}}^{N} {U}/{|\vec r
_i - \vec r_j|} $.  The orbital angular momentum
$M=\sum_{i=1}^{N}m_{(i)}$, the total spin $S$ and $S_z$ (the
projection of the spin along $\hat z$) are good quantum numbers.

By increasing $\vec{B}$, the dot undergoes a sequence of transitions
to higher $M$ and higher $S$ states. These transitions have been
monitored in the conductance for larger dots including tenths of
electrons as well as for dots with few electrons\cite{osterk}.
Eventually, the GS reaches the maximum $M=N(N-1)/2$ and full
spin polarization $ S = N/2 $(MDD) \cite{stopa,noi}. 

The confinement of the QD in the $\hat z $ direction produces an
electric field orthogonal to the dot plane, which gives rise to a
Rashba term in the Hamiltonian, that couples orbital and spin
dynamics.  In a biased dot the size of this perturbation would also
depend on the screening of the source drain bias $V_{sd} $ applied to
the contacts.  The single particle hamiltonian now reads:
\begin{eqnarray}
H_{(i)}\rightarrow H_{(i)}+ \frac{\alpha}{\hbar} \left( \hat {z} \times
\left(\pp_i+\frac{e}{c}\vec A _i\right) \right)
\cdot \vec{\sigma_i}\: .
\end{eqnarray}
Here $\vec{\sigma}$ are the Pauli matrices and $\alpha$ (measured in
units of meV \AA ) is the SO coupling parameter, which is proportional
to the electric field in the $\hat z $ direction.  Good quantum
numbers labeling the multiparticle states are now $ N,S,J_z, E $,
where $J_z=M+S_z$, the total angular momentum along $z$ and $E$ is the
energy.  Details of our exact diagonalization procedure have been
reported previously
\cite{jouault}.  

SO coupling lifts the spin degeneracy of the $M$ multiplets, and the
FSP GS attains the lowest $J_z$: $ J_z^{FSP} = N(N-1)/2 - N/2 $ ($S_z$
is the projection of the spin along $\hat z$).  In a previous work we
exhibited the charge density and spin density of the first excited
state (FES) (see Fig.1,2), which showed unexpected spin texture
properties \cite{noi}. Indeed, the FSP QD reproduces in a nutshell the
Quantum Hall Ferromagnet (QHF) at filling one which has skyrmion-like
low lying collective excitations\cite{sondhi,macdonald,exp}.
\begin{figure}
\includegraphics*[width=\linewidth]{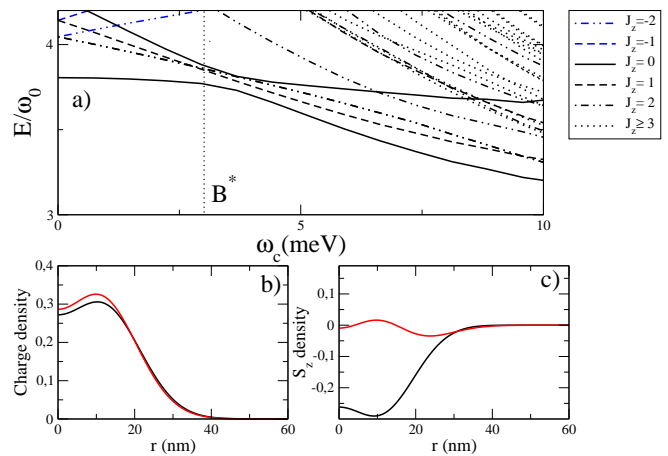}
\caption{(color on-line)
N=2 particles dot: a) energy spectrum vs magnetic field $\omega_c$ in
the presence of the $SO$. Values of the parameters are in the
text. The GS is $J_z=0$, the FES is $J_z=1$. b) Charge densities of
the GS (black line) and of the FES (red line) of the FSP dot.  c)
Corresponding spin densities of the GS (black line) and of the FES
(red line) of the FSP dot.}
\label{plot2}
\end{figure}
\begin{figure}
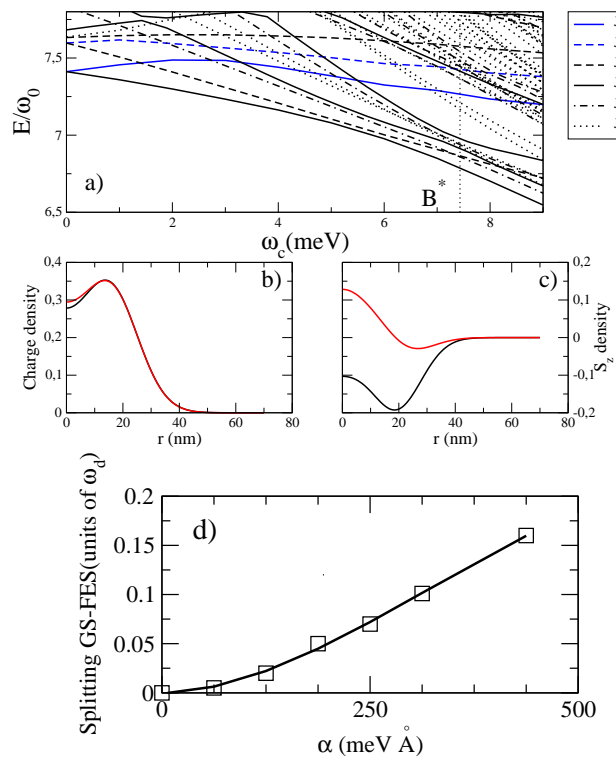

\includegraphics*[width=\linewidth]{ene3.eps}
\includegraphics*[width=0.8\linewidth]{plotvsa.eps}
\caption{(color on-line)
N=3 particles dot: a) energy spectrum vs magnetic field $\omega_c$ in
the presence of the $SO$. $\omega_d=7meV$, $U=13meV$, $\alpha =
250meV$\AA. The GS is $J_z=3/2$, the FES is $J_z=5/2$. b) Charge
densities of the GS (black line) and of the FES (red line) of the FSP
dot.  c) Corresponding spin densities of the GS (black line) and of
the FES (red line) of the FSP dot.  d) GS-FES spin gap vs $\alpha$ at
$\omega_c=8meV$.}
\label{plot3}
\end{figure}
\begin{figure}
\includegraphics*[width=\linewidth]{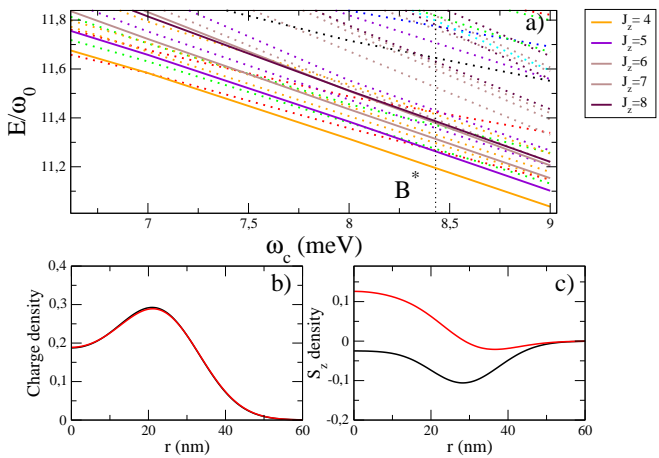}
\caption{(color on-line)
N=4 particles dot: a) energy spectrum vs magnetic field $\omega_c$ in
the presence of the $SO$.  The GS is $J_z=4$, the FES is
$J_z=5$. $\omega_d=7meV$, $U=13meV$, $\alpha = 250meV$\AA. b) Charge
densities of the GS (black line) and of the FES (red line) of the FSP
dot.  c) Corresponding spin densities of the GS (black line) and of
the FES (red line) of the FSP dot.}
\label{plot4}
\end{figure}
\begin{figure}
\includegraphics*[width=\linewidth]{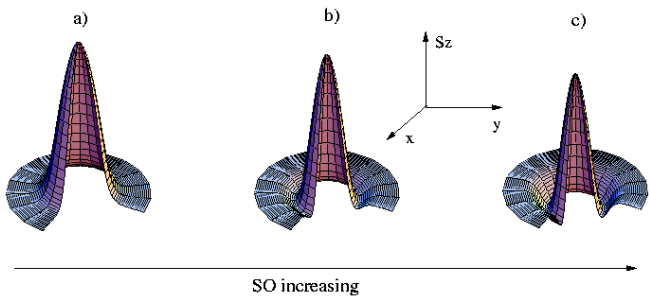}
\caption{(color on-line)
N=3 particles dot: FES spin density (arb.units) for a)
$\alpha=150meV$\AA, b) $\alpha=250meV$\AA, c) $\alpha=350meV$\AA. By
increasing the SO there is a squeezing close to the center and some
reduction of $\langle\sigma_z\rangle$.}
\label{squeez}
\end{figure}
In Fig.1a the lowest lying energy levels $E$ are plotted vs $\omega_c$
for $N=2$ with $\omega _d = 5meV$, $U =13meV$ and $\alpha = 250meV
$\AA.  The level structure is qualitatively analogous to that obtained
in ref.\cite{destefani}, intended for an $InSb$ dot, with Dresselhaus
and cubic SO terms included.  The singlet-triplet transition appears
here as a marked anticrossing at $\hbar \omega_c
\approx 4$meV, because of the SO coupling.  The states involved in the
anticrossing have $J_z = 0 $ and originate, in the absence of SO, from
the singlet $( S=0,S_z =0, M=0 )$ and the triplet $(S=1,S_z=-1,M=1 )$
states. Recently, the relaxation time $T_1$ for the flipping of the
two-electron spin trapped in a vertical $GaAs $ QD, from the triplet
to the singlet state, has been measured, by applying electrical pulses
to the QD.  $T_1$ has been estimated to be $ > 200 \:\mu s $ at $T <
0.5 K $\cite{fujisawa}.  Similarly to what found in
ref.\cite{destefani}, exchange interaction produces a small zero-field
splitting between the first excited state (a triplet with $J_z =2 $ )
and the second excited state (a singlet with $J_z =1$ ).  Increasing
$B$ further the SO induces the crossing of the latter two states, so
that lowest lying states are the GS $ (S =1, J_z =0) $ and the FES $
(S=1, J_z = 1)$.  This crossing qualifies $B^*$ which is rather
insensitive to SO coupling.

As seen from figure\label{plot3,plot4}, the same pattern can be found
also for $N=3,4$.  The SO coupling tends to shift the $\uparrow$ spin
density w.r.to the $\downarrow$ one radially\cite{noi}.  The shift can
occur easily for the GS when $N=2$ and provides a reduction of the
$e-e$ interaction by leaving an isolated spin at the center of the
dot.  When $N>2$, the confinement potential together with the $e-e$
repulsion contrasts such a spin redistribution and the final result is
that the $z-$component of the total spin density is diminished at the
center of the dot. In particular, $\sigma _z (r)$ tends to flatten in
the GS for $N=3,4$.
Correspondingly the radial component $\sigma_r(r)$ increases in the
case of $N=3,4 $ at any distance from the center and not only at the
dot boundary as it happens for $N=2$.

Anticrossings are less prominent for $N=4$ and the level separation of
the bunch of states in Fig.\ref{plot4}a) is much smaller, but a gap
develops at $\omega_c \approx 8.5 meV$, between the GS $ (S =2, J_z
=4) $ and the FES $(S=1, J_z = 5)$. The gap is strongly sensitive to
the SO tuning and increases with increasing $\alpha$ (see
Fig.\ref{plot3}d for $N=3$).

In the FES the SO enforces a spin texture with $\langle S_z \rangle$
flipped at the origin with respect to the GS and healing back
gradually away from the center up to the QD boundary, where the spin
density points radially in the dot plane \cite{noi}. The FES for $B >
B^*$ has $J_z$ increased by one w.r. to the GS. This is mostly due to
spin reversal because the difference of the angular momentum
expectation values $  \langle M
\rangle_{FES} -\langle M\rangle_{GS} $ is found to be vanishingly
small.  In a disk shaped dot, a radial change of $ \langle M
\rangle $ requires a change of $n(r)$ as well, but, as a matter of
fact, we find that the charge distribution in the dot at the FSP point
is rather insensitive to excitation and to the strength of the SO
coupling (see fig.s\ref{plot2}b), \ref{plot3}b),
\ref{plot4}b)).  While the radial charge density $n(r) $ appears to be
compressible at fields $B<B^*$ and $B>B^*$, it is approximately
incompressible at $B\sim B^* $. When $B>>B^*$, the charge distribution
of the dot reconstructs \cite{reconstruction,gudmundsson,noi}.

The spin excitation gives rise to an extra collective magnetization $
\hat z \cdot
\Delta {\vec{\cal{M}}}(r) \approx \langle2 \mu_B
\Delta  \sigma _z (r ) \rangle $,
where $\Delta {\sigma_z } ( r ')$ is the difference in $z$-component of
the local spin density between the FES and the GS and $ \mu _B = e
\hbar / 2 m_e c $.  The radial spin density $\sigma _z (r) $ appears
in Fig.\ref{plot2}c), Fig.\ref{plot3}c),
Fig.\ref{plot4}c) for  $N=2,3,4$ respectively).

We have estimated the possible extra magnetic flux $\phi$ associated
to the spin excitation, by integrating numerically the vector
potential, induced by the spin polarization of the dot, $ a_\vartheta
(r) $, along the circle of radius $R$ at the dot boundary $\gamma$ (
$\phi = \int _\gamma R\: d\vartheta \: a_\vartheta (R) $).  This is
given by:
\begin{equation}
 a_\vartheta  (r)  =   \int _0 ^{2\pi} d\vartheta '\:
\int_{0}^{R} \frac{  dr' \; \vec{ r}' \times \hat z }
{|\vec{ r} -\vec{ r} ' |}\; \frac{\partial  \Delta {\cal{M}}_z
 ( r ')}{\partial r'}\:\:   ,
\label{pot}
\end{equation} 
where $\hat r$ is a radial unit vector. The calculation yields a
fraction of the flux quantum $\sim 10^{-5}
hc/e$, but it is remarkable that, at $B \approx B^* $, we find the
same value of $\phi $ for $N =2,3,4$. This is consistent with the fact
that the FES has essentially one spin flipped at the origin and no
change in orbital angular momentum.

\textit{FIR  absorption}.
Far infrared radiation is a common tool in large scale QD arrays (e.g.
$In$ QD's\cite{fricke} or field-effect confined $GaAs$
QD\cite{krahne}). In the presence of a Rashba SO term the center of
mass coordinate and the relative coordinates are coupled
together\cite{jacak}, so that Kohn's theorem does not apply.  It
follows that FIR radiation could excite the many body FES.  We have
calculated the dipole matrix element squared for the transition from
GS to FES vs B.  Our results are shown in Fig. \ref{ap} for $N=2$(a)
and $N=4$(b), respectively.  The dispersion of the absorption peaks is
artificial, but their detailed shape would yield direct access to the
coupling constants and to the relaxation mechanisms.  We find an
increase of the expected intensity at the FSP point which marks the
crossover to the new states. As expected, the crossover sharpens with
increasing $N$.

\textit{Conclusions}.
Simultaneous application of an electric field with a magnetic field
orthogonal to a disk shaped QD reproduces the properties of a 2DEG QHF
on the dot scale, with its skyrmion excitations. SO opens an
anticrossing gap above the GS and stabilizes a skyrmion like FES,
within the gap, with a spin reversed at the origin of the QD.  FIR can
excite the dot thus affecting nuclear magnetic resonance of underlying
nuclear spins, as in recent experiments in $GaAs$ quantum wells
\cite{salis}.

Discussions with S.Tarucha are gratefully acknowledged, as well as
hospitality at ICTP, Trieste.
\begin{figure}
\includegraphics*[width=\linewidth]{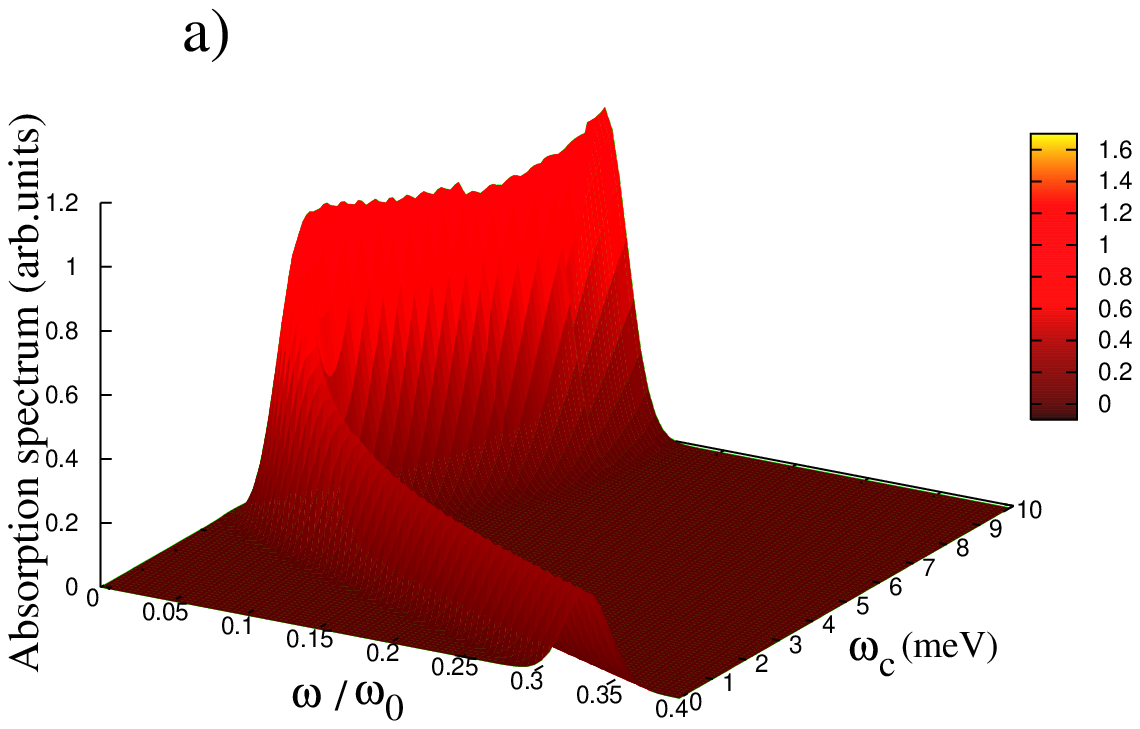}
\includegraphics*[width=\linewidth]{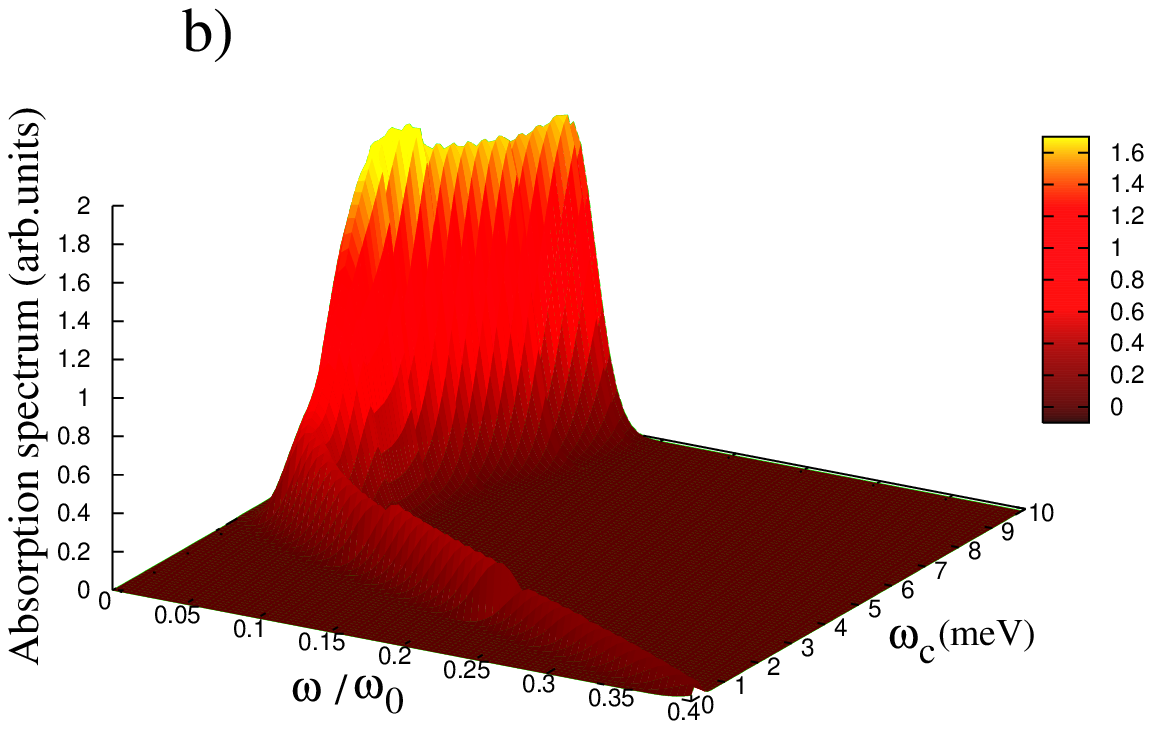}
\caption{(color on-line) 
          Absorption spectrum vs magnetic field for 2(top),4(bottom)
          particles}
\label{ap}
\end{figure}

\end{document}